\documentstyle[preprint,pre,aps,epsf,graphicx]{revtex}
\begin{document}

\draft

\title{Mononuclear caloric curve in a mean field model}
 
\author{S. Das Gupta}

\address{Physics Department, McGill University, 
Montr{\'e}al, Canada H3A 2T8}

\date{\today}

\maketitle

\begin{abstract}
A mean field model is used to investigate if a plateau in caloric curve
can be reached in mononuclear configuration.  In the model the configuration
will break up into many pieces as the plateau is approached.
\end{abstract}


The appearence of a plateau in the experimental nuclear caloric curve,
since it was pointed out first in \cite{Pochodzalla}, has continued to be
an important issue in intermediate energy heavy ion collisions
\cite{Dasgupta1}. The plateau would signify a maximum in specific heat 
and could be a signature of a phase transition. 

The appearence of a maximum in the specific heat at temperature 
$T\approx$ 5 MeV
was seen in theoretical calculations much earlier \cite{Bondorf1}.  The
model is the SMM (Statistical Multifragmentation Model) where it is
assumed that the nucleus breaks up into many pieces at a volume
significantly larger than the normal nuclear volume.  Subsequent modelling
has reinforced this picture of a heated nucleus breaking up into
pieces as the phenomenon which gives a peak in the specific heat.
Take, for example, the LGM (Lattice Gas Model) \cite{Pan1}.  Here at
low temperature there is a large percolating cluster.  As the temperature
is raised one reaches a point when the percolating cluster breaks up into
many smaller clusters.  It is here that the maximum in specific heat is
seen (Fig.(19.2) in \cite{Dasgupta1}).  This is more dramatically
demonstrated in the canoncal thermodynamic model (which is in the same
spirit as the SMM but is much easier to implement).  Here also at low
temperature there is a large blob of matter which breaks up into many 
pieces and again the maximum in specific heat is obtained in this 
region.  The thermodynamic model can be extrapolated to large numbers
of particles.  In this limit it is shown that the break up is very 
sudden as a function of temperature.  This is a case of first order
phase transition and the maximum of specific heat is obtained at the
phase transition.

The present work is inspired by a recent calculation which showed
that, in a different model, a plateau in the caloric curve (hence
a maximum in specific heat) can be reached in mononuclear configurations
as well \cite{Sobotka1}.  One assumes that $e*$(=the excitation energy
per particle) as a function of density
increases quadratically about the ground state density.  At a given 
excitation energy, the nucleus expands in a self-similar fashion
till it reaches its maximum entropy.  Effects of interaction
on entropy is taken through a parametrisation of $m*/m$ where
$m*$ is the effective mass.  
A temperature is defined microcanonically.  The authors then find
that when they plot temperature against excitation energy a plateau
is found around 5 to 6 MeV.

A more familiar model in nuclear physics which allows study of 
the caloric curve in mononuclear configurations is the temparature
dependent mean field model (Hartee-Fock and/or Thomas-Fermi model).
This has certain advantages.  When one does a standard mean field 
calculation at a fixed temparture, one minimises the free energy
$F=E-TS$ \cite{March}.  This means that when we get the self-consistent
solution at a given temperature, we have obtained a solution which
has zero pressure.  If we draw a caloric curve with energies of these
solutions this caloric curve pertains to zero pressure.  The specific heat
that we will get will be $c_p$ with $p=0$.

Investigation of the caloric curve with temperature-dependent Thomas-Fermi
theory was done in the past \cite{De}.  For nuclei $^{150}$Sm and
$^{85}$Kr caloric curves were drawn and
a maximum in specific heat at temperature 10 MeV was found.  But 
there is ambiguity whether the systems stay mononuclear or not.  Even
though we are interested in finite nuclei, when we do finite temperature
mean field theory beyond a certain temperature, fermi occupation
factors for orbitals in the continuum will cause the finite system
to spread out.  The calculations have to be done in a big box and
the size of the box will affect the answers.  Although given the
box, definite numerical results can be obtained it will be hard to decide
whether it is a mononucleus or a system of gas.  This is best explained
using figs. 1 and 2 of \cite{De}.  Looking at figures one would venture
to say that at temperature 5 MeV we have a mononucleus but at temperature
10 MeV the nucleus spreads to 12.5 fm with density $\approx 0.008 fm^{-3}$.
Is this a gas or a nuclear liquid?

A much more definite answer can be obtained in the nuclear matter limit.
This is pursued in this work.  For a fixed temperature we 
do calculations for different densities.  The density where the free energy
per particle is minimised is the solution for this temperature; $e*$
of this solution is the appropriate $e*$ for this temperature. 
As expected, starting from zero temperature, the system expands.
The minima of free energy drop to lower and lower density as the
temperature increases.  But beyond a certain temperature, minimum in free 
energy disappears.  The nucleus will now break apart.  This happens
before $T$ flattens out as a function of $e*$.  This is shown in figs. 1
and 2.

It remains to give some details of the calculation.  We use the 
mometum dependent mean field of \cite{Welke,Gale1}.  The potential
energy density is given by
\begin{eqnarray}
v(\rho)=\frac{A}{2}\frac{\rho^2}{\rho_0}+\frac{B}{\sigma+1}\frac{\rho^{\sigma+
1}}{\rho_0^{\sigma}}+\frac{C}{\rho_0}\int\int d^3pd^3p'\frac{f(\vec r,\vec p)
f(\vec r,\vec p')}{1+[\frac{\vec p-\vec p'}{\Lambda}]^2}
\end{eqnarray}
Here $f(\vec r,\vec p)$ is phase-space density.  In nuclear matter, at 
zero temperature $f(\vec r,\vec p)=\frac{4}{h^3}\Theta (p_F-p)$
where 4 takes care of spin-isospin degeneracy.  At finite temperature
the theta function $\Theta(p_F-p)$ is replaced by Fermi occupation
factor (see details below). The potential felt by a particle is
\begin{eqnarray}
u(\rho,\vec p)=A[\frac{\rho}{\rho_0}]+B[\frac{\rho}{\rho_0}]^{\sigma}
+2\frac{C}{\rho_0}\int d^3p'\frac{f(\vec r,\vec p')}{1+[\frac{\vec p-
\vec p'}{\Lambda}]^2}
\end{eqnarray}
Here $A=$-110.44 Mev, $B$=140.9 MeV, $C$=-64.95 MeV, $\rho_0=0.16 fm^{-3}$,
$\sigma$=1.24 and $\Lambda=1.58p_F^0$.  This gives in nuclear matter
binding energy per particle=16 MeV, saturation density $\rho_0=.16 fm^{-3}$,
compressibility $K$=215 MeV and $m*/m$=.67 at the fermi energy;
$u(\rho,p)$ gives the correct general behaviour of the real part
of the optical potential as a function of incident energy.  A
comparison of $u(\rho,p)$ with that derived from UV14+UVII potential
in cold nuclear matter can be found in \cite{Gale1}.  The specific
functional form of the momentum dependent part arises from the Fock
term of an Yukawa potential.   Mean fields given by eqs. (1)
and (2) have been widely 
tested for flow data \cite{Zhang} and give very good agreement.

To do a finite temperature calculation the following steps have to be
executed.  We need to find the occupation probability
\begin{eqnarray}
n[\epsilon(p)]=\frac{1}{e^{\beta[\epsilon(p)-\mu]}+1}
\end{eqnarray}
for a given temperature $1/\beta$ and density $\rho$.  If $\epsilon(p)$
were known {\it a priori}, this would merely entail finding the chemical
potential from
\begin{eqnarray}
\rho=\frac{16\pi }{h^3}\int_0^{\infty} p^2n[\epsilon(p)]dp
\end{eqnarray}
But the expression for $\epsilon(p)$ is
\begin{eqnarray}
\epsilon(p)=\frac{p^2}{2m}+A[\frac{\rho}{\rho_0}]+B[\frac{\rho}{\rho_0}]^
{\sigma}+R(\rho,p)
\end{eqnarray}
where at finite temperature
\begin{eqnarray}
R(\rho,p)=2\frac{C}{\rho_0}\frac{4}{h^3}\int d^3p'n[\epsilon(p')]\times
\frac{1}{1+[\frac{\vec p-\vec p'}{\Lambda}]^2}
\end{eqnarray}
Thus knowing $R(\rho,p)$ requires knowing $n[\epsilon(p')]$ 
already for all values of
$p'$.  This self-consistency condition can be fulfilled by an iterative
procedure (details can be found in \cite{Gale1}).

To calculate pressure, we use the thermodynamic identity $pV=-E+TS+\mu N$
which then gives
\begin{eqnarray}
p=a+b+c 
\end{eqnarray}
Here
\begin{eqnarray}
a=-v-\frac{16\pi}{h^3}\int_0^{\infty}dp\frac{p^4}{2m}n[\epsilon(p)]
\end{eqnarray}
where $v$ is given by eq. (1) and the second term is the contribution from
the kinetic energy.
\begin{eqnarray}
b=-T\frac{16\pi}{h^3}\int_0^{\infty}p^2[n\ln n+(1-n)\ln (1-n)]dp
\end{eqnarray}
The term $c$ is $\mu\rho$.  The free energy per particle is $-(a+b)/\rho$.
The test $p=0$ when the minimum of free energy is reached provides
a sensible test of numerical accuracy.

A second set of calculation was done with a potential energy density
without any momentum dependence.  This means the $C$ terms in eqs. (1) and (2)
are put to zero and $A,B$ and $\sigma$ are readjusted to give desired values
of binding energy per nucleon, equilibrium density and compressibility.
The constants now are $A$=-356.8 MeV, $B$=303.9 MeV and $\sigma=$7/6.
We study equilibrium values of density with temperature with both
the interactions but so, as not to clutter up figures, only the curves with
momentum dependent interaction is shown in fig.1.  In fig.2 we draw the caloric
curves with both sets of interaction and for reference, Fermi gas
results $e*=aT^2$ are also shown.

In \cite{Sobotka1}, the effects of interactions on the caloric curve were taken
through an effective mass $m*$ replacing the real nucleon mass $m$. Normally
one writes $m*/m=m_km_{\omega}$.  In momentum independent mean field
both $m_k$ and $m_w$ are ignored and $m*/m$=1.  In the momentum dependent
mean field calculation that we have done here $m_k$ is in but $m_{\omega}$
is not.  But in both the models, the nucleus will break up before
$T$ flattens out against $e*$.  As in \cite{Sobotka1}, with momentum
dependence the curve is initially above the Fermi gas and meets the
Fermi gas curve at some later point. 

The mean field model thus reinforces what has long been perceived
in other models.  At low temperature, the system has a large mass.
There are probably some lighter masses too.  As the temperature increases
the larger mass which has the bulk of the system will break up.
This is the point where the caloric curve flattens out.  We also note
that experimentally the interesting features of the caloric curve
occur around 6 MeV temperature whereas in figs 1 and 2 they are
occurring around 12 MeV temperature.  The finiteness of the system
will bring this down as will the Coulomb interaction.  In any case
some overestimation in mean field theory is expected.  In the Ising
Model mean field calculations overestimate the critical temperature 
by fifty per cent \cite{Huang}.

Lastly the lessons from the simple mean field model done here can be 
used to bolster the expanding emitting source (EES) model of W. Friedman
\cite{Friedman}.
The model assumes that initially the hot system evaporates as well
as expands.  For low initial temperature the system will cease expanding 
and will revert towards normal density.  But beyond a certain temperature
at the end of this slow expansion the system will explode.

I thank Lee Sobotka for many communications.  This work is supported
in part by the Natural Sciences and Engineering Research Council of Canada
and in part by the Quebec Department of Education.

\begin{figure}
\epsfxsize=5.5in
\epsfysize=7.0in
\centerline{\epsffile{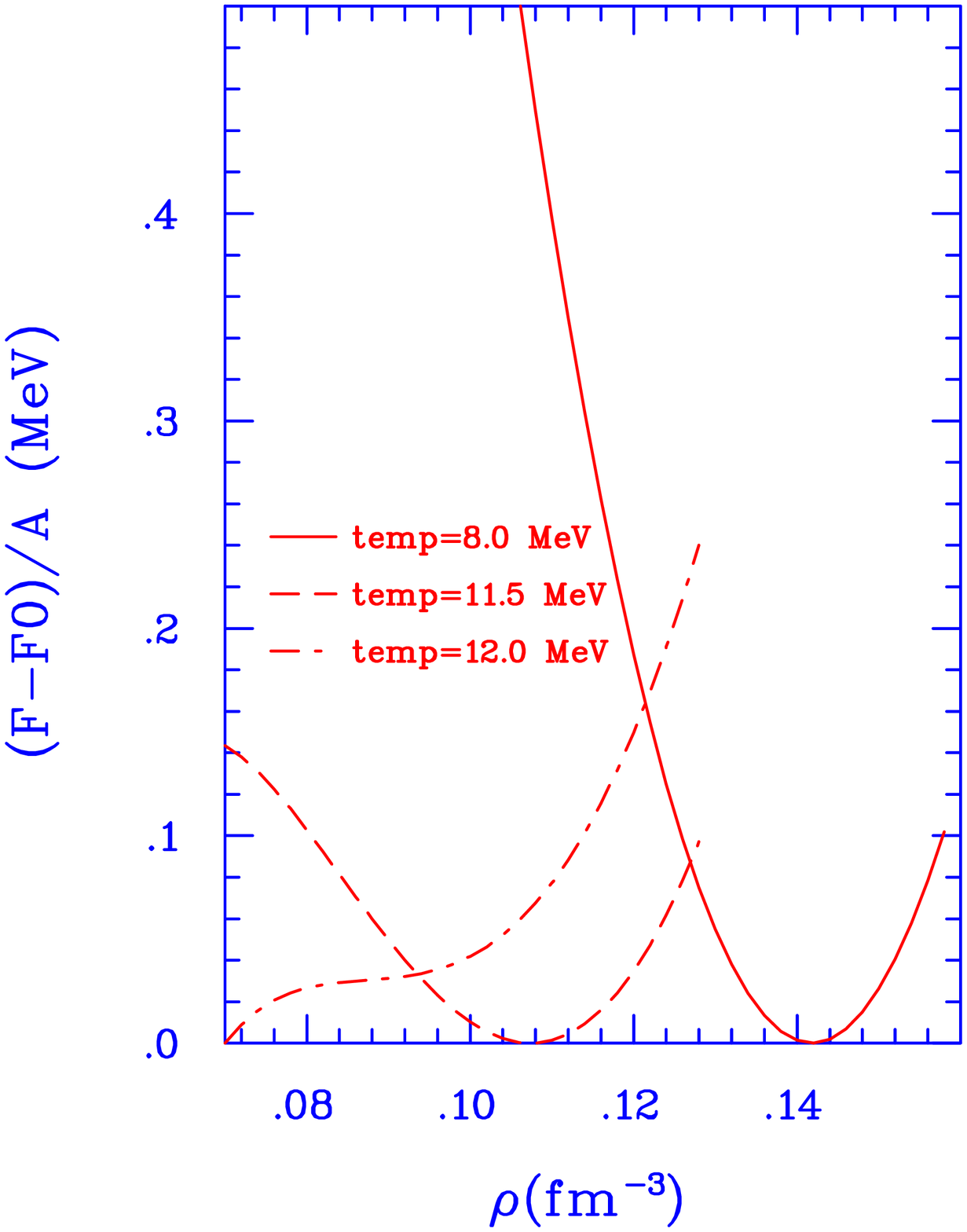}}
\vskip 0.8 true cm
\caption{The free energy per particle as function of density at different
temperatures.  This is plotted as a difference from the minimum value of
F in the frame. At temperature 12 MeV there is no minimum and the system
will roll down to lower density.  For this temperature  already at 
lower values of $\rho$ in the figure, the 
derivative $\frac{\partial p}{\partial\rho}$ is negative, i.e.,
the system has entered a region of mechanical instability.} 
\end{figure}

\begin{figure}
\epsfxsize=4.5in
\epsfysize=6.0in
\centerline{\epsffile{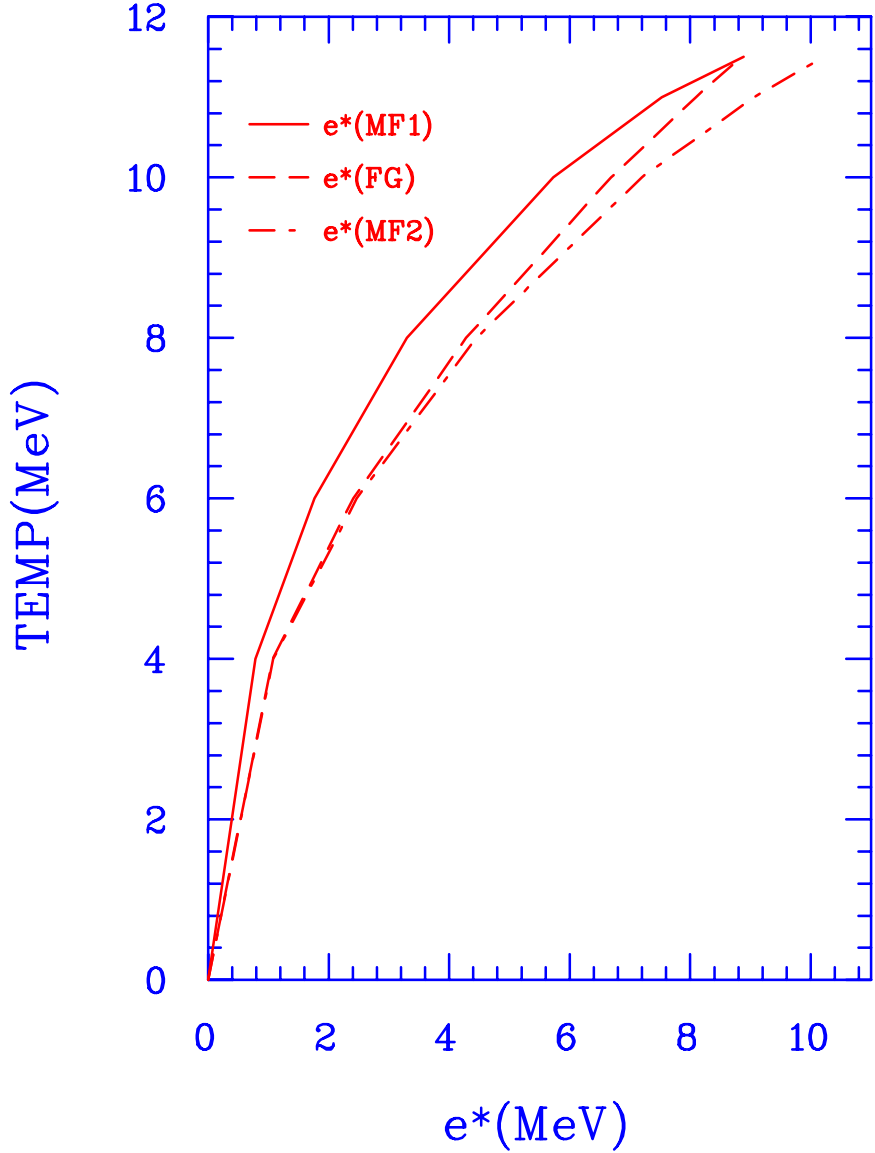}}
\vskip 0.8 true cm
\caption{Temperature against e* in mean field model with momentum
dependence (MF1), in Fermi gas model (FG) and in mean field model without
momentum dependence (MF2).  As in ref. 5, the curve with
momentum dependence lies above the Fermi gas curve till they meet
(around temperature 11.5 MeV).  The MF1 curve has not yet flattened out.
Just below temperature 12 MeV the curve stops.  The nucleus will break
up into many pieces at higher temperature.  Similar situation happens
with MF2 except that here it will break up many pieces at $\approx$ 12.5 
temperature again before the caloric curve flattens out.}
\end{figure}

\end{document}